\documentclass[aps,paper,nofootinbib, superscriptaddress]{revtex4}

\usepackage[english]{babel}
\usepackage[T1]{fontenc}
\usepackage{amsmath}
\usepackage{latexsym}
\usepackage{dsfont}
\usepackage{amsfonts}
\usepackage{graphicx}

\newcommand{\be}{\begin{eqnarray}}
\newcommand{\ee}{\end{eqnarray}}
\newcommand{\bite}{\begin{itemize}}
\newcommand{\eite}{\end{itemize}}

\begin{document}
\title{Computing the Effective Hamiltonian of Low-Energy Vacuum Gauge Fields}
\author{R. Millo\footnote{{\it Current Address:}\\ Theoretical Physics Division, Department of Mathematical Sciences\\ University of Liverpool\\Liverpool, L69 3BX, UK}}
\affiliation{Universit\`a degli Studi di Trento, Via Sommarive 14, Povo (Trento), Italy.}
\affiliation{INFN, Gruppo Collegato di Trento, Via Sommarive 14, Povo (Trento), Italy.}
\author{P. Faccioli}
\affiliation{Universit\`a degli Studi di Trento, Via Sommarive 14, Povo (Trento), Italy.}
\affiliation{INFN, Gruppo Collegato di Trento, Via Sommarive 14, Povo (Trento), Italy.}
\begin{abstract}
A standard approach to investigate the non-perturbative QCD dynamics is through vacuum models which emphasize the role played  
by specific gauge field fluctuations, such as instantons, monopoles or vortexes.
The effective Hamiltonian describing the dynamics of the low-energy degrees of freedom in such approaches is usually postulated phenomenologically, or obtained through 
uncontrolled approximations.  
In a recent paper, we have shown how lattice field theory simulations can be used to 
 rigorously compute the effective Hamiltonian of arbitrary vacuum  models by  stochastically performing the path integral over all the vacuum field fluctuations which are not explicitly taken into
 account.
In this work, we present the first illustrative application of such an approach to a gauge theory and we use it to compute the instanton size distribution in $SU(2)$ gluon-dynamics
 in a fully model independent and parameter-free way.   
\end{abstract}
\maketitle
\section{Introduction}
Contemporary lattice gauge theory  (LGT) simulations allow to compute from first principles a large class of hadronic matrix elements in QCD, in some cases even within a few percent accuracy. 
On the other hand, such simulations do not provide much detailed information about the structure of the gluonic fluctuations which drive the QCD dynamics
 in the strongly coupled regime. 
For example, despite several decades of investigations,  the dynamical processes underlying chiral symmetry breaking and color confinement are still a matter of
debate.

The problem of identifying the dynamical origin of such non-perturbative phenomena has been extensively addressed in the context of phenomenological models which emphasize 
the role played by specific  vacuum field fluctuations,  such as e.g. instantons~\cite{instantonrev}, monopololes~\cite{monopolerev} and  center vortexes~\cite{vortexesrev}. 
The configuration space of these models is defined by the collective coordinates of the selected low-energy vacuum fields. 
On the other hand, the statistical distribution of such collective coordinates (or, equivalently, their effective Hamiltonian) is usually obtained through approximations upon which one does not
have full theoretical control, e.g.  by completely neglecting the contribution of the fluctuations around the chosen vacuum fields~\cite{NegLenz} or by estimating the role of such fluctuations through 
variational methods~\cite{diakonov84}.

In principle, LGT simulations can be used to test the predictions of the phenomenological vacuum models,  for example by looking for some specific signatures of the dynamics generated by 
instantons \cite{myPRL, gattringer}, monopoles \cite{monopolelat} or vortexes \cite{vortexlat}.  
On the other hand,  the model dependence associated to the effective Hamiltonian for the vacuum field degrees of freedom makes it is difficult to draw definitive conclusions about the validity of a given model. 
Indeed, a moderate disagreement with the experimental data 
or with the results of lattice QCD simulations may be due to either a wrong choice of the low-energy vacuum fields, or to the strong approximations involved in the definition of their partition function.

In order to tackle this problem, in a recent work we have developed a technique, which we shall refer to as Vacuum Manifold Projection (VMP), by which lattice 
simulations are used to rigorously compute the 
effective Hamiltonian of arbitrary vacuum models~\cite{paper1}, in a model independent way. This is done by non-perturbatively performing the path integral over all the vacuum 
field configurations which are not explicitly taken into account in the given low-energy vacuum model. For example, in an instanton model one performs the path integral over all the configurations which are orthogonal to the functional manifold spanned by multi-instanton configurations.  
Clearly, once the partition function has been evaluated from first principles, any failure of the model must be entirely due to the wrong 
choice of the effective low-energy degrees of freedom. 

In our first work,  the VMP method was illustrated and tested by evaluating the instanton-antinstanton interaction in a simple quantum-mechanical toy model~\cite{paper1}. Here, we present the first application to a gauge theory. 
In particular, we use the VMP method to compute the instanton size distribution in two-color Yang-Mills theory. 
 
 The paper is organized as follows. In section \ref{method} we review the VMP method for a generic choice of vacuum field degrees of freedom. In section \ref{ILM}
 we focus on an effective theory based on instanton degrees of freedom and we use the VMP method  to compute the instanton size distribution in $SU(2)$ gluon-dynamics. 
The main results, conclusions and perspectives are summarized in section \ref{conclusions}.    

 \section{The Vacuum Manifold Projection Method}
 \label{method}
 
 Let us consider a gauge theory defined by the (Euclidean) path integral
\be
\label{rep1}
Z = \int~\mathcal{D} A_\mu~ e^{-S[A_\mu]},
\ee
where the $S[A_\mu]$ formally includes the gauge-fixing and ghost terms, along with the fermionic determinant. In the following, we shall always assume that this path integral is defined in a Landau gauge.

Let $\gamma\equiv(\gamma_1, \ldots, \gamma_k)$ be a set of $k$ collective coordinates which parametrize a manifold $\mathcal{M}$
of vacuum field configurations $\tilde A_\mu(x; \gamma_1, \ldots, \gamma_k)$. For example, in  instanton models,  $\gamma_1, \ldots, \gamma_N$ are the  positions, sizes and color orientations of all the pseudo-particles in the instanton 
ensemble.  However, in general, we do \emph{not} require the field configurations  $\tilde A_{\mu}(x;\gamma_1, \ldots, \gamma_k)$  to be solutions of the classical Yang-Mills equations of motion.  

For any given choice of the set of collective coordinates $\gamma$,  a generic vacuum gauge field configuration $A_\mu(x)$ contributing to the path integral~(\ref{rep1}) can be decomposed as
\be
\label{decompose2}
A_\mu(x)\equiv \tilde A_\mu(x;\gamma_1, \ldots, \gamma_k)+B_\mu(x),
\ee
where $B_\mu(x)$ will be referred to as the "fluctuation field".  Our goal is to use LGT to perform the path integral over such a field.
More precisely, we want to compute the function $\mathcal{H}(\gamma_1, \ldots, \gamma_k)$  such that 
\be
\label{rep2}
Z=  \int~\mathcal{D} A_\mu~ e^{-S[A_\mu]} = \int d\gamma_1 \ldots d\gamma_k~ e^{-\mathcal{H}(\gamma_1,\ldots,\gamma_k)}.
\ee
Eq. (\ref{rep2}) defines a statistical model in which $\gamma_1, \ldots, \gamma_k$ are the
effective low-energy degrees of freedom and  $\mathcal{H}(\gamma_1, \ldots, \gamma_k)$ is the effective Hamiltonian. We shall see shortly that such function is defined as the logarithm of a gauge-fixed 
path integral.

\begin{figure}[t]
\includegraphics[width=9cm]{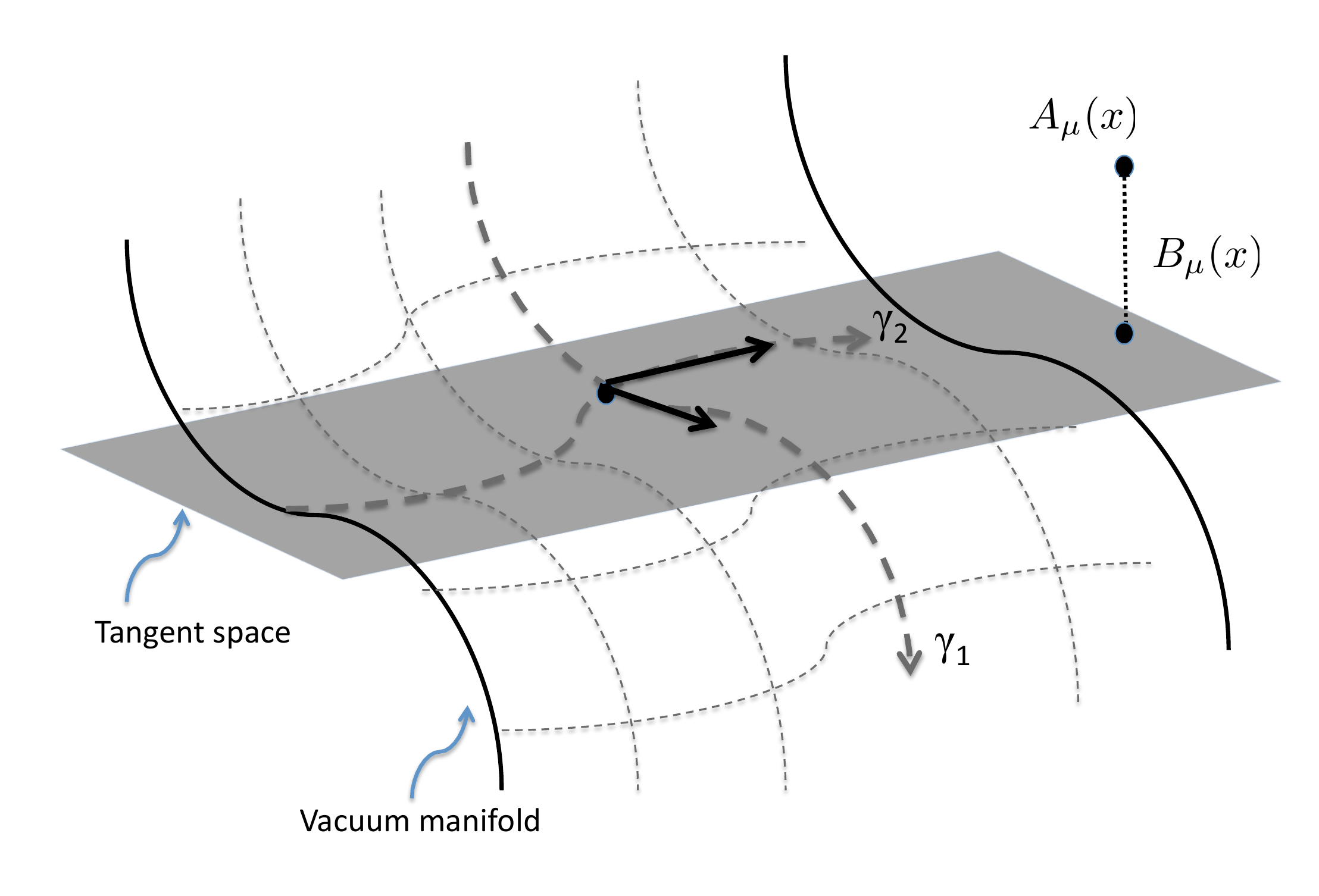}
\caption{Pictorical representation of the projection of the gauge field configuration $A_\mu(t)$ of the path integral $Z= \int \mathcal{D} A_\mu e^{-S[A_\mu]}$  
onto a specific  vacuum field manifold, spanned by two collective coordinates $\gamma_1$, and $\gamma_2$. 
A generic path is represented by a point in this three-dimensional space. The constraints given in Eq.~(\ref{ort2}) imply that the
the fluctuation field $B_\mu(x)$ is perpendicular to the plane tangent to the manifold in a given point.  }
\label{Fig1}
\end{figure}

Since the representation (\ref{rep2}) of the path integral contains $k$ additional integrals over $d\gamma_1 , . . . , d\gamma_k$, we need to introduce $k$ constraints. 
A natural choice is to impose a set of $k$ orthogonality conditions: 
\be
\label{ort2}
\left( B_\mu(x) , g_{\gamma_i, \mu} (x,\bar\gamma) \right) &\equiv& \mbox{Tr}_c\left\{\int\mbox{d}^4 x~ B_\mu(x) ~ g_{\gamma_i,\mu} (x,\bar\gamma)\right\}=0, \qquad i=1,\ldots, k\\
g_{\gamma_i,\mu} (x,\bar\gamma) &=& \left.\frac{\partial}{\partial \gamma_i} \tilde A_\mu(x;\gamma)\right|_{\gamma=\bar \gamma}.
\label{ggamma2}
\ee
We observe that the functions $\{g_{\gamma_i} (x,\bar\gamma)\}_{i=1,\ldots,k}$ identify the $k$ directions tangent to the
 manifold $\mathcal{M}$ of background vacuum fields,  in the point of curvilinear coordinates  $\bar \gamma=(\bar \gamma_1,\ldots,\bar\gamma_k)$ --- see Fig. \ref{Fig1} ---.
We consider only choices of the manifold $\mathcal{M}$ and of the point $\bar \gamma$ such that the vectors (\ref{ggamma2}) define a system of coordinates.

In the path integral formalism, the orthogonality conditions (\ref{ort2}) can be implemented by introducing a Fadeev-Popov representation of the unity. 
After some formal manipulation (see e.g. Ref.~\cite{paper1, diakonov84} )  one arrives to an expression in the form of Eq.~(\ref{rep2}), where the effective Hamiltonian
$\mathcal{H}(\gamma_1, \ldots, \gamma_k) $ is defined as 
 \be
\label{2Gdef2}
\mathcal{H}(\gamma_1, \ldots, \gamma_k)&=&  -\log \bigg\{\int\mathcal{D}B~\delta[\partial_\mu B_\mu]~\prod_i \delta\left[ \left(B_\mu(x),  g_{\gamma_i,\mu} (x, \bar{\gamma})\right)
\right]~\Phi[\tilde{A}(x,\gamma)+B(x)]e^{-S[\tilde{A}(x,\gamma)+B(x)]}\bigg\},
\ee
where $\Phi$ is a Jacobian factor and reads
\be
\label{2Gdef2b}
\Phi^{-1}[A_\mu(x)]&=&\int\prod_{l=1}^k~d\gamma_l\int\mathcal{D}U^\Omega~\delta[\partial_\mu A^\Omega_\mu]~\prod_i 
\delta\left[\left( A^\Omega_\mu(x)-\tilde{A}_\mu(x,\gamma) , g_{\gamma_i,\mu} (x,\bar{\gamma})\right)\right].
\ee
$U^{\Omega}(x)$ denotes a generic gauge transformation and   $A_\mu^\Omega(x)$ is result of gauge transforming the field $A_\mu(x)$ according to $U^{\Omega}(x)$.
 Note that, while the path integral $Z$ is obviously gauge invariant, the definition of the effective Hamiltonian $\mathcal{H}(\gamma_1,\ldots, \gamma_k)$  relies on the choice of a Landau gauge.

The orthogonality condition (\ref{ort2}) and the system of coordinates (\ref{ggamma2}) can be used to device an algorithm to explicitly compute the effective Hamiltonian (\ref{2Gdef2}). 
We begin by  observing that,  on such a system of coordinates, a functional point  $A_\mu(x)$ which belongs to the vacuum field manifold  $\mathcal{M}$ is identified by the coordinates $(\Psi_1,\ldots, \Psi_k)$ with
\be
\label{lpk}
\Psi_i[A_\mu(x)] & = & \left(A_\mu (x) , g_{\gamma_i,\mu} (x,\bar\gamma)\right).
\ee
Clearly, also field configurations which lie in a functional  neighborhood  of the manifold $\mathcal{M}$ can be projected onto the same system of coordinates. In this case,  
for an arbitrary fixed choice of $\bar \gamma$, the orthogonality conditions (\ref{ort2}) imply that there exits a set of collective coordinates $\gamma$ such that
\be
\Psi_i [A_\mu(x)] &\equiv& (A_\mu(x),  g_{\gamma_i, \mu} (x, \bar\gamma) ) \nonumber\\
&=& ( \tilde A_\mu(x;\gamma)+B_\mu(x),  g_{\gamma_i, \mu} (x, \bar\gamma) ) \nonumber\\
&=&
 (\tilde A_\mu (x; \gamma), g_{\gamma_i,\mu} (x, \bar\gamma) ),
 \qquad (i=1, \ldots, k).
\label{ort3}
\ee

Such an equation allows to associate a set of collective coordinate $\gamma=(\gamma_1, \ldots, \gamma_k)$ to any Landau gauge-fixed gluon field configuration $A_\mu(x)$ which lies in the functional 
vicinity of the manifold $\mathcal{M}$. Based on this result,  it is immediate to device an algorithm to evaluate the path integral $\mathcal{H}(\gamma_1, \ldots, \gamma_k)$ stochastically, from LGT simulations:
\begin{enumerate} 
\item An ensemble of $N_{conf}$ independent lattice 
configuations $\{ U^{(1)}_\mu(x),\ldots, 
U^{(N_{conf})}_\mu(x) \}$ is generated by standard LGT simulations.

\item From such  configurations, an ensemble of Landau \emph{gauge}-fixed lattice configurations 
$\{ U^{g~(1)}_\mu(x),\ldots, 
U^{g~(N_{conf})}_\mu(x) \}$ is obtained, e.g. using  the procedure illustrated in  Ref.s~\cite{citeLandau1, citeLandau2}.

 \item From each lattice configuration, the gluon field $A^g_\mu(x)$ at each lattice site is calculated 
from the gauge-fixed lattice link variables $U^g_\mu(x)$, for example using the discretized definition
\be
A^g_\mu(x)=\frac{1}{2ai}\left(U^g_\mu(x)-U^g_\mu(x)^\dagger\right).
\ee
\item 
$k$ non-linear equations for the $\gamma_1,\ldots, \gamma_k$ variables are obtained from Eq.s (\ref{ort3}):
\be
\left\{
\begin{matrix}
(A^g_\mu(x) , g^g_{\gamma_1,\mu} (x,\bar\gamma) ) =
(\tilde A^g_\mu(x, \gamma) , g^g_{\gamma_1,\mu} (x,\bar\gamma) ) \\
...\\
\label{system3}
 (A^g_\mu(x),  g^g_{\gamma_k, \mu} (x, \bar\gamma) )=  (\tilde A^g_\mu (x \gamma), g^g_{\gamma_k,\mu} (x, \bar\gamma) ).
\end{matrix}
\right.
\ee
On the lattice, the inner products above are obviously represented as discretized sums, 
\be
(A^g_\mu(x) , g^g_{\gamma_i,\mu} (x,\bar\gamma) ) = a^4\sum_n\mbox{Tr}_C\left[A^g_\mu(n) g^g_{\gamma_i}(n;\bar\gamma)\right],\qquad (i=1, \ldots, k).
\ee  
where $n$ runs over all the lattice sites and $\textrm{Tr}_C$ refers to the trace over color labels.

Notice that the quantities on the left-hand-side of Eq.s (\ref{system3}) are c-numbers which depend only on the lattice configuration and on the projection point $\bar\gamma$. They correspond to the components of the 
lattice field  $A^g_{\mu}(x)$ in the system of coordinates
defined by the tangent vectors $g^g_{\gamma_l, \mu}(x, \bar \gamma)$.

On the other hand,  the quantities on the right-hand-side of Eq. (\ref{system3}) are functions  of the curvilinear coordinates $\gamma$ which do not depend on the lattice
 configurations $A^g_{\mu}(x)$. 
These functions are completely specified, for any given fixed choice of the vacuum field manifold $\mathcal{M}$ and of the projection point $\bar \gamma$.   

Hence, for any lattice  configuration $U^g_{\mu}(x)$ the set of equations (\ref{system3})  
can be numerically  solved for  $\gamma=(\gamma_1, \ldots, \gamma_k)$.  

\item The frequency histogram of $\gamma_1, \ldots, \gamma_k$ obtained by repeating this procedure for a large number $N_{conf}$ of statistically independent lattice gauge field configurations represents the probability 
$\mathcal{P}(\gamma_1, \ldots,\gamma_k)$ of a given set of curvilinear coordinates. 
Then the effective Hamiltonian for $\gamma_1, \ldots, \gamma_k$ is obviously given by
\be
\mathcal{H}(\gamma_1, \ldots, \gamma_k) = - \log \mathcal{P}(\gamma_1, \ldots, \gamma_k).
\ee
\end{enumerate}

Some comments on the VMP algorithm are in order. First of all, we emphasize that such a scheme relies on the assumption that the 
background field configurations $\tilde A_\mu(x; \gamma)$ are the relevant vacuum degrees of freedom. More precisely, we are requiring that the field configurations which contribute the 
most to the path integral (\ref{rep1}) lie in some functional neighborhood of the manifold $\mathcal{M}$, so that they can be projected onto the system of coordinates defined in Eq. (\ref{ggamma2}). 
Under such a condition, the definition of the effective Hamiltonian $\mathcal{H}(\gamma_1, \ldots, \gamma_k)$ is unique and does not depend on any additional external parameters. 

Once the effective Hamiltonian has been determined,   arbitrary vacuum-to-vacuum matrix elements can be 
approximatively computed by neglecting the contribution of the fluctuation field $B_\mu(x)$ to the operators. For example, if $\hat O(x)$ is a local operator which depends on the gluonic field $A_\mu(x)$, one has
\be
 \langle 0 |  \hat{O}[A_\mu(x)] | 0 \rangle = \frac{1}{Z} \int \mathcal{D} A_\mu O[A_\mu(x) ]~e^{-S[A_\mu]} \simeq \frac{1}{Z} ~\int d\gamma_1 \ldots d\gamma_k~O[\tilde A_\mu(x; \gamma) ]~ 
e^{-\mathcal{H}(\gamma_1,\ldots,\gamma_k)}.
 \ee
Since the effective Hamiltonian is evaluated from first principles, any violation of this identity implies that the fluctuation field $B_\mu$ plays an important role in the observable associated to this matrix elements or 
that there are important fluctuations in the path integral which are very far from the functional manifold $\mathcal{M}$, hence cannot be projected onto the system of coordinates (\ref{ggamma2}).  
In both cases, this would represent an unambiguous signature of the failure of the vacuum field model.  

We also stress that the VMP results do not depend on the choice of the gauge. Indeed, in Ref.~\cite{paper1} it was shown that once the system of Eq.s (\ref{system3}) is satisfied in one gauge, 
it holds also in any other gauge. 

Finally, we note that the VMP procedure is conceptually analog to a technique routinely adopted in classical statistical mechanics to evaluate the potential of mean-force $G(\gamma_1, \ldots, \gamma_k)$
as a function of a set of collective coordinates $\gamma_1,\ldots, \gamma_k$~(see e.g. \cite{MolSim} and references therein). 
In the canonical ensemble, the potential of mean-force (free energy) is defined as 
\be
e^{-\beta G(\gamma_1, \ldots, \gamma_k)} = \int d \Gamma ~\prod_{i=1}^k\delta\left[\gamma_i - f_i(\Gamma)\right]~ e^{-\beta H(\Gamma)}\qquad (\beta= 1/ k_B T),
\ee
where $\Gamma$ is the phase-space variable, $H(\Gamma)$ is the Hamiltonian, and the functions $f_i(\Gamma)$ specify the definition of the macroscopic collective coordinate $\gamma_i$ in terms of the microscopic phase-space variable $\Gamma$. 
The system's partition function reads
\be
Z = \int d\gamma_1\ldots d\gamma_k ~e^{-\beta G(\gamma_1, \ldots, \gamma_k)},
\ee
in complete analogy with Eq.~(\ref{rep2}).
In order to evaluate  $G(\gamma_1, \ldots, \gamma_k)$ one generates and ensemble of statistically independent equilibrium configurations $\{\Gamma_1, \ldots, \Gamma_{N_{conf}}\}$,
by means of Monte Carlo or molecular dynamics simulations,  and evaluates $\gamma_1, \ldots, \gamma_k$ from $\gamma_i = f_i(\Gamma)$, for each of such configurations. In the limit of large number of equilibrium 
configurations $N_{conf}$  the frequency histogram for the values $\gamma_1, \ldots, \gamma_k$ obtained this way yields the equilibrium 
 probability function for the collective coordinates, $\mathcal{P}(\gamma_1, \ldots, \gamma_k)$.  
 The negative of the logarithm of such a probability defines by construction the free energy 
 \be
 -\beta G(\gamma_1, \ldots, \gamma_k) = \log \mathcal{P}(\gamma_1, \ldots, \gamma_k).
 \ee

 \section{Calculation of the Instanton Size Distribution in $SU(2)$ gluon-dynamics.}
 \label{ILM}
 
 As a first illustrative application of the VMP method, in this section we specialize to the case 
 in which the gauge theory is $SU(2)$ gluon-dynamics and the vacuum manifold $\mathcal{M}$ is constructed by superimposing singular-gauge instanton and anti-instanton configurations, i.e. in the so-called instanton vacuum. 
  
 Instanton vacuum models have been successfully used to investigate the dynamics of light hadrons \cite{instantonrev} and the breaking of chiral symmetry\cite{chiralsymmetrybyinstantons} in QCD. 
 It has been shown that this model is able to reproduce the spectrum of the
  observed lowest-lying hadrons~\cite{mass1, mass2, mass3} and of scalar and pseudo-scalar glueballs\cite{glueballs1,glueballs2, glueballs3,glueballs4}, along with the electromagnetic structure of 
 pions and nucleons \cite{em1, em2, em3, em4}. In addition, the intanton liquid model provides an explanation of the  $\Delta I=1/2$ rule for non-leptonic hadron decays of
  kaons~\cite{ventodelta12} and hyperons~\cite{ourdelta12},  by promoting non-perturbative scalar diquark correlations~\cite{diquarks}.
  While it is well known that semi-classically inspired vacuum models based only on singular-gauge instantons do not provide color confinement, a finite string tension was obtained in models employing
  cocktails of regular- and singular- gauge 
  instantons and merons \cite{NegLenz, merons}.
 
From the theoretical point of view, a main limitation of the instanton approach to QCD resides in the well known "infrared catastrophe" of the semi-classical dilute instanton gas approximation: in the presence of
quantum fluctuations, isolated instantons tend to swell.  
The instanton liquid model originates from the observation that such an infrared divergence can be cured
 if correlations between pseudo-particles are included~\cite{shuryak82, diakonov84}.  Unfortunately 
allowing for such  interactions implies giving up a rigorous semi-classical theory of the QCD vacuum. In the so-called instanton liquid models the size distribution of the pseudo-particles 
is estimated using  variational or phenomenological arguments. In such approaches, the typical instanton size in QCD is found  to be of the order of $1/3$~fm --- see e.g. \cite{diakonov84, instantonrev}---.  

In order to extract information about the structure of instanton ensemble directly from LGT simulations,  algorithms such as cooling~\cite{cooling} or eigenvalue filtering~\cite{gattringer}
have been proposed. Such methods provide techniques to filter  out ( not integrate out) the high-frequency quantum content of lattice configurations. 
A problem with these methods is that their results
critically depend on the choice of additional uncontrolled parameters, such as the number of cooling steps or of the number of retained low-lying eigen-modes of the Dirac operator. 
Clearly, such a dependence introduces some degree of arbitrariness in the results. For example, the instanton density vanishes if the limit of very large number of cooling steps. 

The goal of this section is to show that the VMP method can be used to rigorously evaluate the instanton size distribution directly from lattice simulations, without introducing any arbitrary parameter. 
The choice of focusing on $SU(2)$ Yang-Mills theory was made in order to keep the analytical and numerical calculations as simple as 
possible.

\subsection{The Instanton Vacuum Manifold}

We begin our calculation by defining the manifold of vacuum field configurations.   In two-color Yang-Mills theory, the classical field of an individual instanton or anti-instanton is specified by four collective coordinates: $\gamma= (z,\rho, 
\theta_1,\theta_2,\theta_3)$ 
where $z$ is the position of the pseudo-particle, $\rho$ is its size, and $\theta_i$ ($i=1,2,3$) are three angles which specify a $SU(2)$ matrix in color space according to:
\be 
U= \exp(i \theta_k \tau_k).
\ee

A configuration of $N_+$ instanton and $N_-$ instantons $A_\mu(x;\gamma_1,....,\gamma_{N_++N_-})$ can be constructed in the so-called sum \emph{ansatz}, i.e. by superposing the 
classical fields of the pseudo-particles:
\be
\label{sumansatz}
\tilde A_\mu(x;\gamma)= \sum_{i=1}^{N_+ + N_-}~\tilde A^i_\mu(x;\gamma_i),
\ee
where $\tilde A^i_{\mu}(x;\gamma_i)$ is the classical field of the $i$-th pseudo-particle:
\be
\label{AfieldsI}
\tilde A^i_\mu(x;\gamma_i)  &=& U^i\tau^aU^{\dagger,i}\bar\eta^{a}_{\phantom{a}\mu\nu}\frac{\rho^2_i}{(x-z^i)^2}\frac{(x-z^i)_\nu}{(x-z^i)^2+\rho^2_i}, \qquad \textrm{(for instantons)}\\
\label{AfieldsA}
\tilde A^i_\mu(x;\gamma_i)  &=&  U^i\tau^aU^{\dagger,i}\eta^{a}_{\phantom{a}\mu\nu}\frac{\rho^2_i}{(x-z^i)^2}\frac{(x-z^i)_\nu}{(x-z^i)^2+\rho^2_i}, \qquad \textrm{(for anti-instantons)}.
\ee
$\bar\eta^{a}_{\phantom{a}\mu\nu}$ and $\eta^{a}_{\phantom{a}\mu\nu}$ are the so-called 't Hooft indexes. It is important to emphasize that the field obtained from the 
sum ansatz (\ref{sumansatz}) is not in general a solution of Euclidean Yang-Mills equation of motion.

An effective theory for the two-color Yang-Mills theory can be obtained if the path-integral over all the gauge field configurations is replaced by the sum over all 
the configurations of a grand-canonical statistical ensemble of 
singular gauge instantons and antiinstantons:
\begin{equation}
\label{eq:partf}
Z_{YM} =\sum_{N_+, N_-} \frac{1}{N_+!N_-!} e^{i \theta (N_+ - N_-)}~\int\prod_i^{N_+ + N_-} d\gamma_i
~ e^{-\mathcal{H}(\gamma_1,\ldots\gamma_N) }
\end{equation}
where  $\theta$ is the angle associated to strong CP violation and $\mathcal{H}(\gamma_1,\dots, \gamma_N)$  is the effective Hamiltonian, which
 represents the  functional integral over the configurations which do not belong
to the instanton vacuum manifold.  
In the analogy with classical statistical mechanics discussed in section \ref{method}, this term can be interpreted as the ``potential of mean-force'' between the pseudo-particles generated by all other
quantum gauge field fluctuations in the path integral.  
Clearly, if the effective Hamiltonian $\mathcal{H}(\gamma_1,...,\gamma_N)$ is calculated non-perturbativelly from first principles, 
then Eq.(\ref{eq:partf})  provides an exact  representation of the path-integral, which does not rely at all on the semi-classical approximation. 

If the instanton ensemble is not too dense, it is possible to perform a many-body expansion of the effective Hamiltonian $\mathcal{H}(\gamma_1,...,\gamma_N)$: 
\be
\mathcal{H}(\gamma_1,...,\gamma_N)\simeq \sum_i u_1(\gamma_i) + \sum_{i<j} u_2(\gamma_i, \gamma_j) + \sum_{i<j<k} u_3(\gamma_i, \gamma_j,\gamma_k) + \ldots
\label{manybody}
\ee 
The gauge invariance and translational invariance of the vacuum  imply that the one-body term of the $i-$th pseudo-particle, $u_1(\gamma_i)$, is only a function of its size $\rho_i$.
Hence, the function $n(\rho_i) = \exp[-u_1(\gamma_i)]$ is called the (single-) instanton size distribution. 
The divergence of such a term in the semi-classical dilute gas limit gives raise to the  "infrared catastrophe". 
The terms $u_2(\gamma_i, \gamma_j), u_3(\gamma_i, \gamma_j, \gamma_k), \ldots$ describe multi-body interactions,  and depend in general also on the relative positions and color orientations
 of the pseudo-particles.  
With such a definition, and for $\theta=0$ the partition function reads
\be
\label{eq:partf2}
Z_{YM} =\sum_{N_+, N_-} \frac{1}{N_+!N_-!} ~\int\prod_i^{N} d\gamma_i
~\left( \prod_{i=1}^{N_{+} + N_-} n(\rho_i)\right)~e^{-\sum_{i<j} u_2(\gamma_i, \gamma_j) + \sum_{i<j<k} u_3(\gamma_i, \gamma_j,\gamma_k) + \ldots}
\ee
In the following, we present a computation of the single instanton size distribution $n(\rho)$. The calculation of the many-body terms is conceptually analog and is referred to future work.  

\subsection{Computing the instanton size distribution  $n(\rho)$. }

In order to calculate $n(\rho)$, we have implemented the VMP method on the functional manifold spanned by individual instantons (and anti-instantons) defined in the singular gauge, which is a  Landau gauge.  
The system of coordinates $\{ g_{\gamma_i,\mu} (x,\bar\gamma)\}$ was obtained by numerically differentiating the instanton field (\ref{AfieldsI}) and (\ref{AfieldsA})
 with respect to each of the eight collective coordinates: 
\be
 g_{\rho,\mu} (x,\bar\rho, \bar z, \{ \bar \theta_i\}) &=& \frac{\tilde A_\mu(x;\bar \rho+\Delta \rho)-\tilde A_\mu(x;\bar\rho-\Delta \rho)}{2 \Delta\rho},\nonumber\\
 g_{z^\nu,\mu} (x,\bar\rho, \bar z, \{ \bar \theta_i\}) &=& \frac{\tilde A_\mu(x;\bar z^\nu+\Delta z^\nu)-\tilde A_\mu(x;\bar z^\nu-\Delta z^\nu)}{2 \Delta z^\nu},\qquad (\nu=1,2,3,4)\nonumber\\
 g_{\theta_i,\mu} (x,\bar\rho, \bar z, \{ \bar \theta_i\}) &=& \frac{\tilde A_\mu(x;\bar\theta_i+\Delta\theta)-\tilde A_\mu(x;\bar\theta_i-\Delta\theta)}{2 \Delta \theta_i},\qquad (i=1,2,3),
\ee
where $\bar\rho, \bar z^1, \bar z^2, \bar z^3, \bar z^4, \bar \theta_1,\bar \theta_2,\bar \theta_3 $ define an arbitrary point on the manifold, whose choice will be specified below.

In the VMP method, the distribution of collective coordinates is obtained by projecting lattice field theory configurations, and solving the system of equations~(\ref{lpk}). 
We used three different ensembles of $SU(2)$  Landau gauge-fixed  lattice QCD configurations, generated using the Wilson action~\cite{andresternbeck}~\footnote{We thank 
 Andre Sternbeck for providing us with such configurations.}: 
\begin{itemize}
\item an ensemble of 100 configurations generated on a $16^4$ lattice at $\beta=2.3$ (corresponding to a lattice spacing $a\simeq0.17$~fm)
\item an ensemble of 100 configurations generated on a $24^4$ lattice at $\beta=2.4$ (corresponding to a lattice spacing $a\simeq0.12$~fm)
\item an ensemble of 25 configurations generated on a $32^4$ lattice at $\beta=2.5$ (corresponding to a lattice spacing $a\simeq0.09$~fm).
\end{itemize}
The gauge fixing of such configurations was performed using an over-relaxation algorithm --- see \cite{citeLandau1, citeLandau2}---.

For each lattice configuration we have used 10 different projection points $\bar \gamma =(\bar \rho,  \bar z^1, \bar z^2, \bar z^3, \bar z^4, \bar \theta_1, \bar \theta_2,\bar \theta_3)$. The color orientation angles $ \bar \theta_i$ 
and the space-time coordinate $ \bar z^\nu$ of the projection points were chosen randomly. 
On the other hand, the instanton size  $\bar \rho$  of the projection point was held fixed to five lattice spacings, $\bar \rho = 5~a$. This choice was made in order to ensure that $\bar \rho$
was always much larger than the lattice spacing $a$, yet much smaller than the size of the simulation box.

Once the  projection equations (\ref{system3}) were solved numerically for all lattice configurations and for all projection points, the  probabilities for the collective coordinates
  $\rho$, $z^\nu$, $\theta_1, \theta_2$ and $\theta_3$ were inferred from the corresponding frequency histograms.  
The distribution of the positions $z$, and color orientation angles $\theta_1, \theta_2, \theta_3$ are trivial, as a consequence of the gauge invariance and translational invariance of the vacuum. 
On the other hand, the probability distribution $n(\rho)$ for the instanton size $\rho$ was found to display non-trivial structure, as it is expected as a consequence of  dimensional transmutation. 

\begin{figure}[t]
\includegraphics[width=8.5 cm]{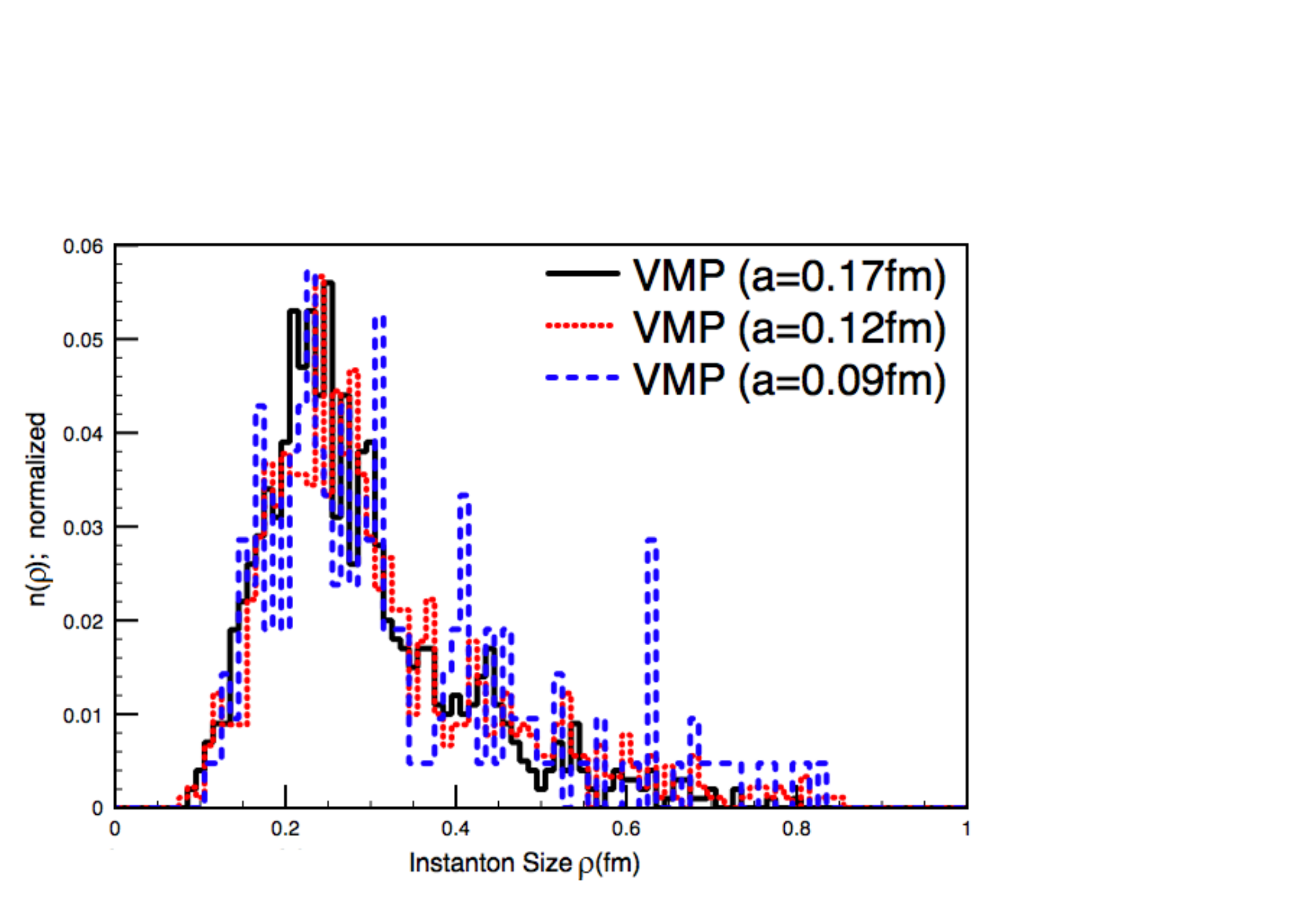}
\includegraphics[width=8.5 cm]{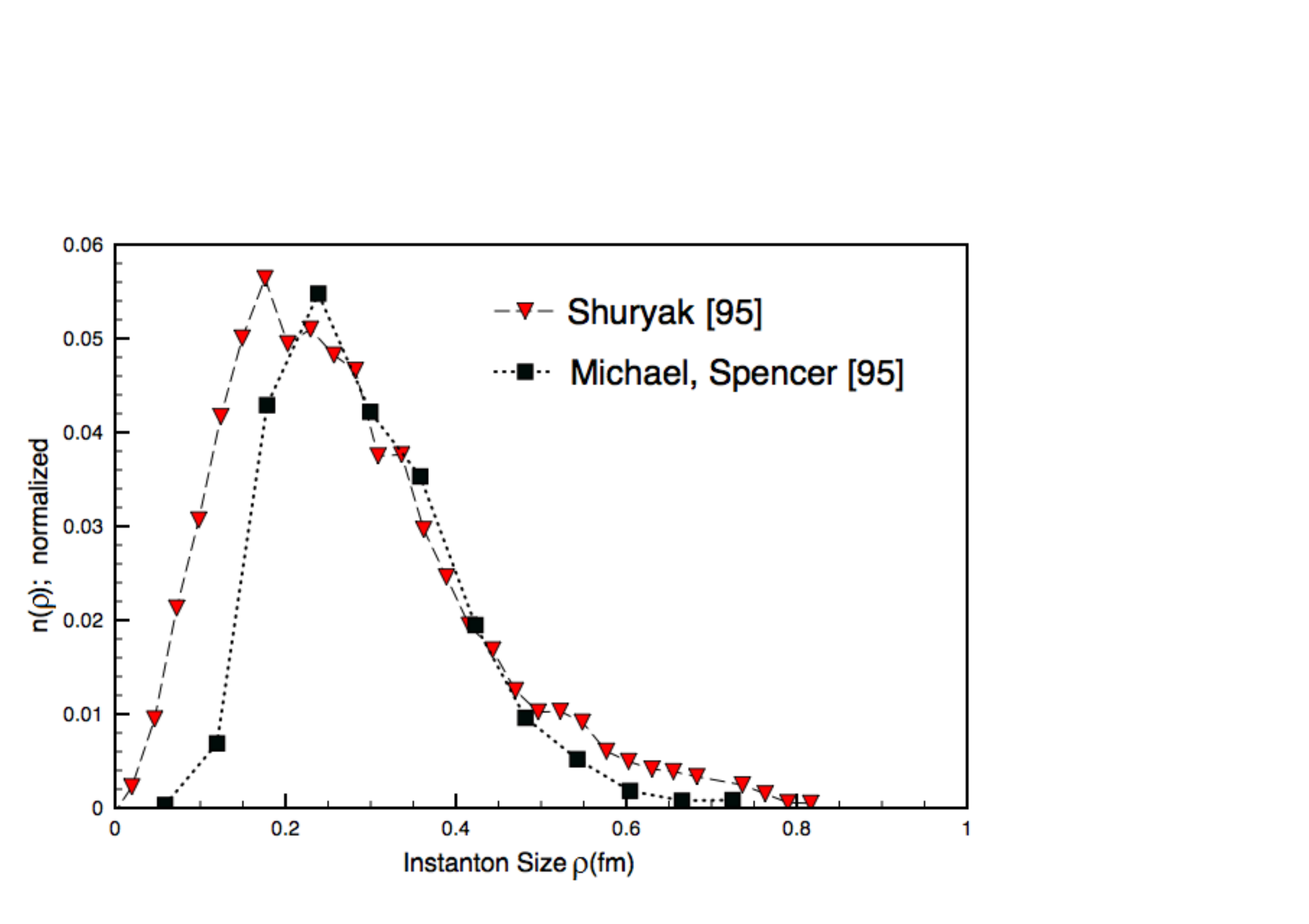}
\caption{Left panel: Instanton size distribution $n(\rho)$ obtained with the VMP technique, from three different sets of {\it gauge-fixed} configurations. 
The results have been obtained with $\bar\rho=5a$.  Right panel: Results for pure gauge $SU(2)$ obtained by C.Michael and S.Spencer from lattice simulations using  the  
cooling algorithm (squares)  \cite{coolingconfront} and by Shuryak from interacting instanton liquid model simulations (triangles) \cite{shuryakconfront}.}
\label{true}
\end{figure}

The instanton size distributions $n(\rho)$ obtained from the three different ensembles of lattice configurations are shown in Fig.~\ref{true}. 
We see that, within the statistical errors, the results are independent from the value of the lattice spacing $a$. The obtained distributions are peaked around $\rho\simeq~0.2~\textrm{fm}$ 
and qualitatively agree  with the 
results of  other methods (see right panel of Fig.~\ref{true}). On the other hand, we stress that the present results do not depend on any arbitrary parameter, and take into account the full quantum  content of the lattice configurations. 

\begin{figure}[t]
\includegraphics[width=9 cm]{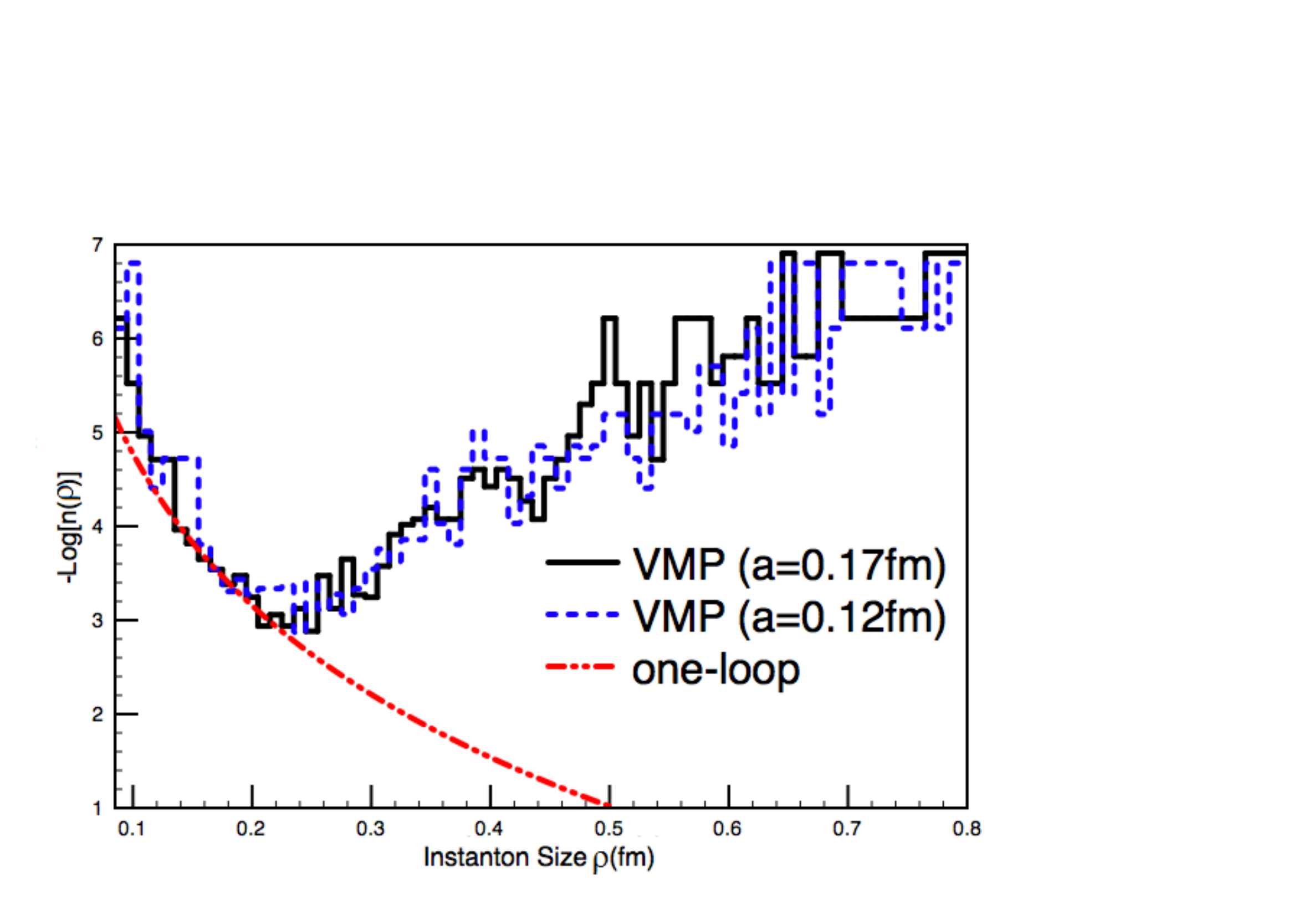}
\caption{Logarithm of the instanton size distribution $n(\rho)$, obtained with the VMP technique from two different sets of {\it gauge-fixed} configurations. The results are compared with the exact one-loop calculation obtained by 't Hooft \cite{thooft}. The VMP results are in agreement with the one-loop calculation, in the small-sized instantons range $0.1~\textrm{fm}\lesssim\rho\lesssim 0.2$~fm. The VMP results
for very small-sized instantons ($\rho \lesssim 0.1$~fm) is affected by lattice discretization errors.  }
\label{pert}
\end{figure}

It is instructive to compare the small instanton size tail of the $n(\rho)$ distribution obtained non-perturbatively through the VMP method 
with the analytical one-loop formula  obtained by 't Hooft \cite{thooft},
\be
n_{one-loop}(\rho)\propto\rho^{\frac{11}{3}N_c-5}.
\ee
In Fig.~\ref{pert} we show that our results are  in quantitative agreement with the leading-order perturbative prediction in the range 
$0.1~\textrm{fm}\lesssim\rho\lesssim 0.2$~fm. The VMP results
for very small-sized instantons ($\rho \lesssim 0.1$~fm) are affected by lattice discretization errors, while the perturbative calculation is not reliable for large instanton sizes,
 $\rho \gtrsim 1/\Lambda_{SU(2)}$. 
Such a comparison shows that  the suppression of the large-sized instantons is a purely non-perturbative effect. 

 \subsection{Testing the accuracy of the VMP calculation of the instanton size}
 
\begin{figure}[t]
\includegraphics[width=9cm]{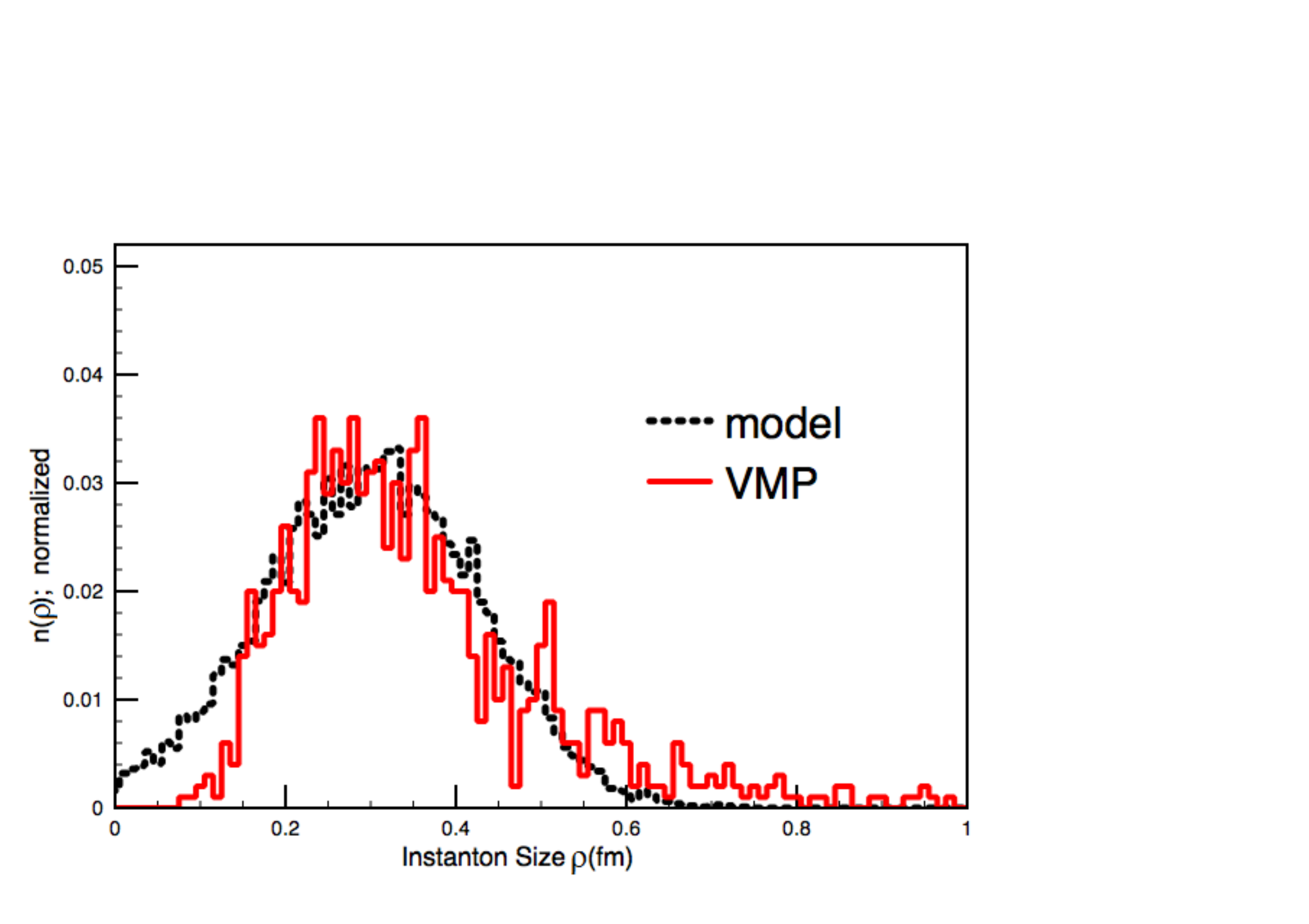}
\caption{Comparison between the distribution used to generate an instanton ensemble with density $n\simeq0.75(fm^{-4})$ and the distribution obtained using the VMP approach from the corresponding 
lattice configurations. }
\label{size1}
\end{figure}

In this section we present a study of the accuracy of our VMP calculation of the instanton size distribution based on an instanton model. We 
generated 1000 configurations of an ensemble of configurations constructed by superimposing the fields of 20 instantons and 20 anti-instantons in a box of volume  $V= (2. 7~\textrm{fm})^4$.  
The positions and color orientations of the pseudo-particles were randomly chosen,  while the size of the pseudo-particles was sampled from a Gaussian distribution. 
\be
n_{\text{model}}(\rho) \propto \exp\left[ - \frac{(\rho-\rho_0)^2}{2 \sigma^2}\right],\qquad (0~\textrm{fm}\le~\rho\le~1~\textrm{fm}),
\ee
with  $\rho_0 = 0.3~\textrm{fm}$, and  $\sigma = 0.13~\textrm{fm}$. The field configurations were discretized on a lattice of spacing $a=0.17$~fm.
The VMP procedure outlined above was applied  to such an ensemble of instanton-model lattice configurations in order to compute instanton size distribution  $n(\rho)$.
In Fig.~\ref{size1} we compare the distribution $n(\rho)$ calculated through the VMP method and the exact $n_{\text{model}}(\rho)$ which was used to generate the instantons ensemble. We see that the VMP approach allows to very accurately reconstruct the correct instanton size distribution for all values of the instanton size larger than about a lattice spacing. 

\section{Conclusions}
\label{conclusions}

In this work we have presented the first application of the recently developed VMP method to a gauge theory.
We have used such a method to perform a non-perturbative calculation of the instanton size distribution in $SU(2)$ gluon-dynamics. Our results are in good agreement
with other lattice calculations for all $\rho$, and with the one-loop perturbative estimate, in the small instanton size regime.  
The VMP calculations of the many-body terms of the effective Hamiltonian of the instanton vacuum models is conceptually analog to the one presented here. 
For example, in order to compute the  instanton-instanton two-body interaction $u_2(\gamma_i, \gamma_k)$ one would need to project onto the manifold spanned by the collective coordinates of two 
pseudo-particles. Such a calculation was performed in \cite{paper1} in the case of  a quantum mechanical toy model and its extension to gauge theories does not raise conceptual difficulties.

It should be emphasized that, in the present  calculation, the fluctuations around the instanton field are not just filtered away, as in the cooling or eigenvalue 
filtering algorithms. Instead, they are systematically integrated out. As a result, the calculation is entirely self-consistent and does not depend on any arbitrary external parameter.    
Recently, Perez and co-workers have proposed a so-called Adjoint Filtering Method based on Dirac quasi-zero modes in the adjoint representation \cite{perez}, 
which does not rely on additional parameters. This method has been used to study the dynamics of smooth single-instanton configurations ''heated'' by Monte Carlo 
updating steps. The present VMP method discussed in this paper could be used in principle to perform the same task. On the other hand, the VMP approach can be applied also
to vacuum field configurations which are not solutions of the equation of motion.

\acknowledgements
P.F. is a member of the Interdisciplinary Laboratory for Computational Science (LISC), a joint venture of Trento University and FBK. R.M. is presently supported by the 
Research Executive Agency (REA) of the European Union under Grant Agreement number PITN-GA-2009-238353 (ITN STRONGnet).
The VMP approach was developed in collaboration with L. Scorzato, who we gratefully acknowledge. We are also indebted with F. Di Renzo for making observations which triggered our work, with F. Pederiva for important discussions, and with A. Sternbeck for providing us with gauge-fixed $SU(2)$ lattice configurations.

\end{document}